\documentclass[twocolumn,prl,amsmath,amssymb,aps]{revtex4}

\begin{document}



\noindent {\bf Comment on "Classifying Novel Phases of Spinor Atoms"}

\vskip 0.5 cm

In Bose Einstein condensates of finite spin atoms,
both spin-rotation
and gauge symmetries can be broken.
In a recent paper, Barnett et al \cite{Barnett06} 
classify these finite spin Bose condensates
 by polyhedra according
to the directions of maximally polarized states
which are orthogonal to the state $\psi$ under consideration.
The purpose of this Comment is to point out that the all
important {\em phase factors} associated with the gauge symmetry
has been left out in \cite{Barnett06}.  In many cases,
the Bose condensate is invariant under a rotation only
when a suitable accompanying gauge transformation is
included.  These phase factors also have non-trivial
consequences in classification of vortices, thus we obtain
results very different from \cite{Barnett06}.

It suffices to illustrate our point by examples.
Consider the ferromagnetic state F of spin $2$,
with state vector \cite{Ciobanu00}
$(1,0,0,0,0)$.  Under $R_z(\alpha)$, a rotation
about $\hat z$ by angle $\alpha$, the state vector
becomes $(e^{ - 2 i \alpha}, 0, 0, 0, 0)$ and 
thus {\it not} invariant (in constrast to \cite{Barnett06}).
The state-vector is invariant only under the combined
operation $R_z(\alpha) e^{2 i \alpha}$.

Next, 
consider the state vector 
$(i, 0, \sqrt{2}, 0, i)$ belonging to the state
named "cyclic" in \cite{Ciobanu00}
(named "tetrahedric" in \cite{Barnett06}, who also
chose a different state vector.  These
state vectors are related by rotation and gauge transformation, as
this state is unique \cite{Mermin74})
To find the symmetry, it is convenient to note that
a state $(\zeta_2, ..., \zeta_{-2})$ has the same rotational
symmetry as the spatial wavefunction
$\psi = \sum_m \zeta_m Y_2^m(\hat k)$, where $Y_l^m$ are
the spherical harmonics.
We then find (ignoring overall real proportionality
constants irrelevant for discussions here and below)
$\psi = \epsilon \hat k_x^2 + \epsilon^2 \hat k_y^2 + \hat k_z^2$
(same form as in \cite{Mermin74}), 
where $\epsilon \equiv e^{2 i \pi/3}$.
It is easy to
see that this state is invariant under two-fold rotations
about $\hat x$, $\hat y$ or $\hat z$.   Under $2 \pi/3$ rotation
about diagonals of a cube, e.g., $(\hat x + \hat y + \hat z)/\sqrt{3}$,
where $(\hat k_x, \hat k_y, \hat k_z ) \to (\hat k_y, \hat k_z, \hat k_x)$,
the state acquires extra phase factors.  Thus the 
isotropy group of the state is \cite{VG,Makela03} 
$\{ E, 3 C_2, 4 C_3 \epsilon, 4 C_3^2 \epsilon^2 \}$
(named ${\rm T(D_2)}$ in \cite{VG})
 Here $E$ is the identity,
and the first $4C_3$ are $2 \pi/3$ rotations about
 $(\pm \hat x \pm \hat y + \hat z)/\sqrt{3}$ or
 $(\mp \hat x \pm \hat y - \hat z)/\sqrt{3}$.
Note the phase factors $\epsilon$'s, which were left out
in \cite{Barnett06}.  
(If we also consider time-reversal symmetry $\Theta$, then the isotropy
group becomes the larger group ${\rm O(D_2)}$:
see \cite{VG}).

For a third example, consider the state A of spin $3$ in
\cite{Diener06}, with state vector $(1,0,0,0,0,0,1)$.
Under $R_z(\alpha)$ the state becomes
$(e^{- 3 i \alpha},0,0,0,0,0, e^{3 i \alpha})
= e^{ - 3 i \alpha} (1, 0, 0, 0, 0, 0, e^{ 6 i \alpha})$.  
Thus under $R_z(\frac{2\pi}{6})$, the
{\it relative} phase between the $m = \pm 3$ components is
unchanged, but the state 
acquires an extra factor $e^{ - i \pi}$.
Hence the invariant operation is $C_6 e^{ i \pi}$ (not $C_6$).
It follows also that the state is invariant
under $C_3 = (C_6 e^{ i \pi})^2$ 
etc.
To find other symmetry operations, we use again the analogy
to $l=3$.  The wavefunction becomes
$- (\hat k_x + i \hat k_y)^3 + (\hat k_x - i \hat k_y)^3$.
It is evident that the state is invariant under $\pi$ rotation
about $\hat y$, and also under $\pi$ rotation about $\hat x$
except a phase factor $e^{ i \pi}$.  The existence of
$C_3$ tells us that there are two other
horizontal two-fold axis $2 \pi/3$ with respect to each of these.
The isotropy group for this state is thus
$\{E, 2 C_3, 2 C_6 e^{ i \pi}, C_2 e^{ i \pi}, 3 U_2 e^{ i \pi},
  3 U'_2 \}$.
(named ${\rm D_6 (D_3) }$ in \cite{VG}).
Note the phase
factors $e^{i \pi}$ accompanying $C_6$, $C_2$ and $U_2$'s,
whereas \cite{Barnett06} simply represented this
state as a hexagon.

These phase factors in the isotropy groups
have non-trivial consequences when classifying vortices.
As an example, 
for the cyclic state, 
vortices are divided into \cite{Makela03,Yip06} 
{\em seven}
classes in additional to the circulation numbers $n$.
(In \cite{Barnett06} however, it was stated that the number of
topological excitations is six, and
circulation numbers were left out.)
We note further that, for
the vortices where the order parameter is rotated by $C_3$ [$C_3^2$]
when one travels along a path encircling that vortex,
the associated phase changes  should be
(see the elements in ${\rm T(D_2)}$)
$( 2 n + \frac{2}{3}) \pi$
[ $ (2 n + \frac{4}{3} ) \pi$],
not the ordinary $ 2 n \pi$. 
These phase factors must be kept correctly to
properly discuss combination of two vortices \cite{Makela03}.
For example, combining two vortices with 
circulations $( 2 n_1 + \frac{2}{3}) \pi$
and $( 2 n_2 + \frac{4}{3}) \pi$
leads to a total circulation of $ 2 (n_1 + n_2 + 1 ) \pi$
but not $ 2 (n_1 + n_2)  \pi$.

In conclusion, we have pointed out that \cite{Barnett06}
has left out phase factors in their discussions on
symmetries and vortices of spin condensates.
More discussions on symmetries of these states can be
found in \cite{Yip06}.

\vskip 0.5 cm
\noindent S.-K. Yip \\
 Institute of Physics, Academia Sinica,\\
 Nankang 115, Taipei, Taiwan \\
 \date{\today} \\
 PACS numbers: 03.75.Hh,03.75.Mn




\end{document}